# Precise control of high-frequency ultrasounds in thin crystals for the development of tunable narrowband and directional γ-ray sources

Emmanouil Kaniolakis Kaloudis[1,2] Nektarios Papadogiannis[1,2,*], Yannis Orphanos[1,2], Makis Bakarezos[1,2] and Konstantinos Kaleris[1,2,**]

[1] *Institute of Plasma Physics and Lasers (IPPL), Hellenic Mediterranean University, GR-74100, Greece*

[2] *Physical Acoustics and Optoacoustics Laboratory, Department of Music Technology and Acoustics, Hellenic Mediterranean University, GR-74100, Greece*

Correspondence to: *npapadogiannis@hmu.gr, **kkaleris@hmu.gr

This work presents a complete methodology for the precise characterization of the acoustic field inside crystal-based devices driven by high-frequency ultrasounds towards the generation of tunable narrowband and directional γ-radiation via undulation of ultra-relativistic charged particles. Such γ-ray sources have long been anticipated by the scientific community, as they promise new powerful tools for the study of high-energy physical phenomena and the development of novel nuclear technologies. In such devices, a piezoelectric transducer induces tens of MHz harmonic waves inside a Silicon monocrystal. Ultra-relativistic charged particles traversing the crystal get trapped within the channels formed by the extremely strong electric fields of the acoustically modulated lattice planes, undergoing undulation and emitting γ-radiation. Precise characterization of the acoustic field in the crystal is crucial for the determination of the expected characteristics of the secondarily generated γ-rays. For this purpose, fast laser refraction imaging is used here to image the acoustic waves by exploiting the spatial redistribution of a laser beam's optical intensity, caused by the acoustic field. A dedicated computational model is developed for the estimation of spatial distribution of the pressure and lattice deformation inside the crystal. This methodology provides a framework for future novel γ-ray sources in high-energy facilities.

# I.    INTRODUCTION

High-frequency ultrasounds (HFUS) in the 10s-of-MHz range lie at the forefront of modern acoustics, thanks to their distinct acoustic properties, and mainly their short wavelength, relatively long penetration depth, high acoustic energy and strong directionality[1-4]. These properties render them suitable for a diverse range of scientific and technological applications, such as high resolution imaging, medical diagnostics and treatment, drug delivery and targeted therapy[5-11]. HFUS are also employed in the development of acousto-optic (AO) devices for laser radiation control, such as modulators (AOMs), beam deflectors and tunable filters[12,13,14].

Recently, HFUS have been proposed for the development of acoustically-driven crystalline undulators (A-CUs) for the generation of tunable narrowband γ radiation[15,16,17]. Gamma rays are electromagnetic radiation at the upper end of the electromagnetic spectrum, characterized by extremely high photon energies and ultrashort wavelengths. They are commonly produced in nuclear reactions, radioactive decay, Bremsstrahlung and Compton scattering (ICS) phenomena[18-21]. They are of critical importance in fundamental physics research, particularly photonuclear physics, probing of nuclear structures, nuclear waste neutralization and medical imaging and treatment[22-25]. State-of-the-art γ-ray sources constitute linear electron accelerators (LINACs) and laser electron accelerators[26,27], Compton scattering sources[28-31], synchrotrons[24] and radioactive isotope sources[32]. However, these technologies have significant restrictions, especially in terms of controllability, tunability and directionality. Compton and synchrotron sources exhibit low efficiency[33], generally resulting to low photon yields, while radioactive isotopes, although capable of generating narrowband γ-rays, offer limited control regarding the intensity and photon energy[34]. Furthermore, γ radiation from Bremsstrahlung processes is neither narrowband nor tunable. These aspects render the aforementioned technologies unsuitable for prospective applications where monochromaticity, tunability and directionality is required.

Another technology worth mentioning is the Free Electron Laser (FEL), which is capable of generating strongly directional, monochromatic and coherent high energy light beams in the X-ray range. In FELs, strong electromagnets are exploited to form a periodic harmonic magnetic field. Ultra-relativistic charged particle beams propagating along the magnetic field follow sinusoidal trajectories in a phenomenon known as undulation. Acceleration of the charged particles leads to the emission of strongly directional, monochromatic and coherent X-ray photons. However, currently available FELs cannot operate beyond the hard X-ray regime and no planned FEL facility is expected to operate in the γ-ray regime. Despite that, FELs are very large and costly and have complex infrastructure and high operational requirements[24,35,36]. Nevertheless, as shown here, their principle of operation opens up the way for the design of novel undulation-based γ sources.

Development of practical, cost efficient, tunable and narrowband γ-ray sources is highly anticipated by the scientific community[37]; however, it has been proven challenging so far. In this direction, Crystal Light Sources (CLS) is a promising technology for the generation of γ-radiation that exploits the extremely strong electric fields formed in the planes of crystal lattices[24,35,36]. CLS technology is currently the main task of the TECHNO-CLS project of the Pathfinder program of the European Innovation Council (EIC)[38]. In CLSs, charged particles, e.g. positrons or electrons, penetrating a crystal in a specific angle can get trapped in the electric fields of the channels formed by the crystal lattice planes. Under the influence of such electric forces, a charged particle accelerates and emits γ-radiation with characteristics that depend on the particle trajectory and kinetic energy[39].

Acoustically-driven Crystalline Undulators (A-CUs) is one of the most promising CLS schemes as it could deliver narrowband and strongly directional γ-rays with tunable photon energy and adjustable brilliance[15,16,17]. In A-CUs, dynamic modulation -bending- of the lattice plains is achieved by longitudinal HFUS travelling within the crystal structure. The HFUS are excited by a multi-MHz piezoelectric transducer bonded to the crystal, i.e. Si, Ge or any other crystal that favors particle channeling. Control of the photon energy and brilliance can be achieved in A-CUs by tuning the frequency and amplitude of the acoustic waves, respectively. Alternative methods for bending the crystal lattice plains, such as periodic mechanical stress, grooving or epitaxially grown superlattices[24,40,41] have limitations in real-time tunability and controllability.

In this work, we present a complete methodology for the design, development and characterization of A-CU devices, in terms of the dynamic acoustic modulation, suitable for undulation of ultra-relativistic positrons. Emphasis is given on the experimental characterization of the acoustic field inside the crystal, which is achieved here via fast laser refraction imaging (FLRI). FLRI exploits the deflection of nanosecond laser pulses due to the periodic gradients of the crystal's refractive index perpendicularly to the laser k-vector. The initial homogeneous optical density of the crystal is dynamically modulated by travelling tens-of-MHz acoustic waves induced by the piezo transducer, which create regions of compression and rarefaction that travel with the speed of sound across the crystal. The snapshot of the induced refractive index periodicity forces the nanosecond laser light to refract, generating an inhomogeneous map of the light intensity field on an imaging screen. A-CU crystals need to be sufficiently thin in order for the charged particles to remain trapped in the lattice channels across their propagation in the crystal, a strict condition for efficient γ radiation emission depending on the charged particle's ultrarelativistic energy. Hence, the optical path of the diagnostic laser beam is typically very short ($\leq$ 1 mm), dictating for a very high sensitivity of the diagnostic method. Also, the dynamic nature of the travelling waves requires fast probing in order to have a static representation of the acoustic field.

Here, experimental results are presented from the characterization of travelling acoustic waves with 10 MHz and 20 MHz frequency in Si crystals with 1 mm thickness. The used prototype devices are suitable for experiments of γ-radiation generation by undulation of ultra-relativistic positron beams with energies of several 10s of GeV, which are available in for example in CERN. Complementarily, a numerical method based on computational modeling and signal processing is developed for the quantification of the acoustic parameters, i.e. pressure amplitude and frequency, which are decisive for the charged particle trajectories and the corresponding resulting γ-ray spectrum. For the initial calibration and validation of the aforementioned computational model, a laser acousto-optic device is used. The calibrated model is then used to quantify the acoustic pressure inside the crystals of the prototype devices. To the best of our knowledge, this is the first demonstration of a complete experimental and computational procedure for the characterization of prototype A-CU devices in terms of the appropriate acoustic fields, to be used in future experiments of γ radiation generation in high energy facilities.

## II.    METHODOLOGY

### A. Experiments

It is well known that when a homogeneous collimated laser beam propagates inside a medium that exhibits refractive index gradients in a direction perpendicular to the laser propagation, it undergoes refraction. Deflection of the light rays due to refraction leads to the deformation of the laser beam profile, which is imprinted as an inhomogeneity in the distribution of the laser intensity[42-45]. A schematic diagram of the abovementioned process is given in FIG. 1., where the light rays (red arrows) undergo refraction from the sinusoidal pressure gradient along the $y$ direction of a crystal with $L_y \times w$ dimensions. This results in a characteristic intensity distribution (blue shaded area) along the $ys$ direction of an imaging screen at distance $L_x$ from the crystal. Under conditions explained later in the manuscript, this intensity redistribution directly reflects the variations in the refractive index inside the medium. As shown later, in case of modulation of the refractive index due to acoustic waves, the light intensity pattern can be used to estimate the acoustic field with high accuracy.

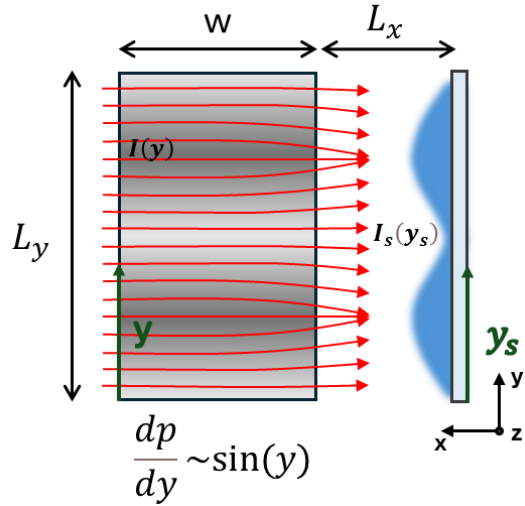

FIG. 1. Schematic diagram of the refraction of the laser field from a crystal with an inhomogeneous refractive index gradient. The red arrows represent the rays of the laser light, which undergo refraction resulting in an inhomogeneous intensity distribution on an imaging screen (blue shaded area). The distance of the imaging screen from the crystal and the height and width of the crystal are denoted as $w, L_x, L_y$, respectively. The refractive index gradient is proportional to the derivative of the pressure along the y axis.

Here, fast laser refraction imaging (FLRI) is achieved by using a pulsed 6 ns Nd: YAG laser system (Quantel Brilliant B) with 1064 nm wavelength and up to 10 Hz repetition rate. Nanosecond laser pulses allow for quasi-static imaging of the travelling acoustic waves. Indicatively, within the 6 ns of the laser pulse, the ultrasonic waves travel by ~50 µm inside the Si crystal, which is smaller than the expected acoustic wavelength by almost an order of magnitude. The schematic diagram of the experimental setup is shown in FIG. 2. The energy of the polarized laser beam is controlled via two successive polarizers. A 10× beam expansion is achieved via two 75 mm and 750 mm converging lenses. Moreover, a 20 µm pinhole is placed on the focal spot of the 75 mm lens to clean the beam from high order diffraction components, thus optimizing the laser facefront. Finally, the light propagates through the sample and is directed onto a CCD camera (Daheng MER2). In the default setup, the probed crystal is placed 30 cm away from the camera sensor. To synchronize image capturing, the CCD camera is triggered by the Q-switch output of the laser. For every measurement, "signal" and "reference" images are captured, corresponding to the laser beam profile after propagating through the crystal with and without acoustic excitation, respectively.

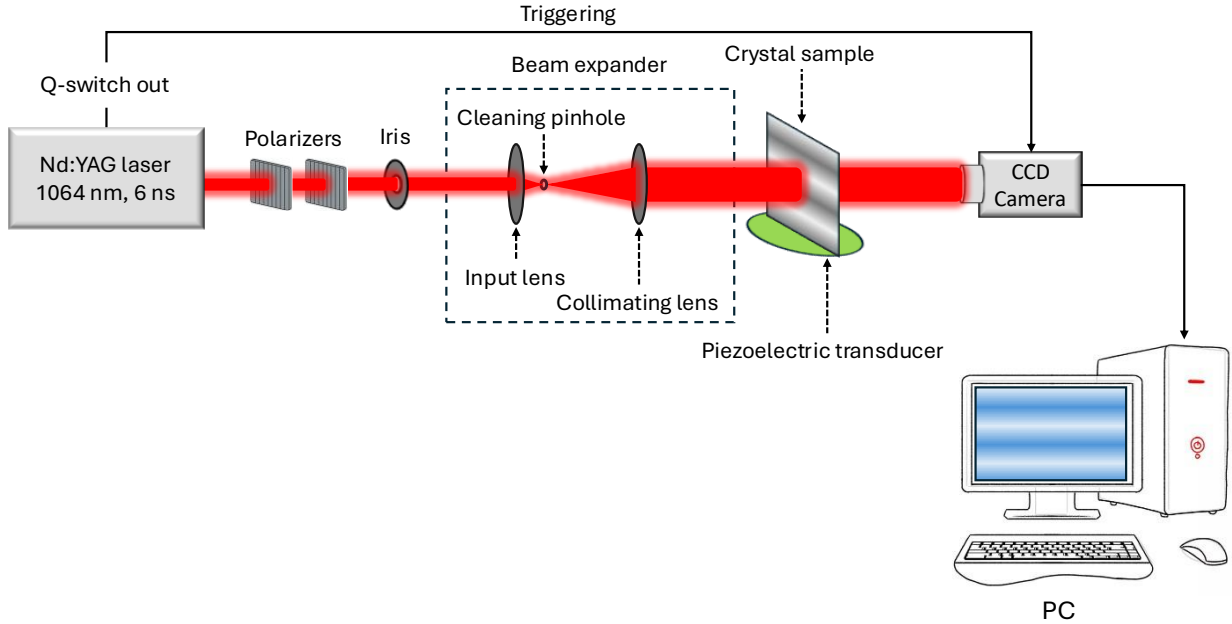

FIG. 2. Schematic diagram of the nanosecond refractive imaging setup used for the detection and characterization of travelling AWs in crystals. A set of polarizers controls the laser beam energy, while an iris is used to select the zero diffraction order. The beam is expanded 10 times by a beam expander consisting of a 75 mm and a 750 mm lens. A 20 μm pinhole placed at the focal spot of the 75 mm lens is used to filter out higher order diffraction components. The cleaned laser beam propagates through the crystal and is captured by a CCD camera.

## B. A-CU prototypes

For the purposes of this study, two prototype devices are developed based on 20 x 20 mm$^2$ and 1 mm thickness Si <100> crystals of ultra-high quality, as specified by the low dislocation density <300 cm$^-$ [2]. Two different piezoelectric transducers with 10 MHz and 20 MHz nominal resonant frequencies are used to excite ultrasonic waves of approximately 800 µm and 400 µm wavelengths, respectively. The transducers are made of lead zitronate titanate (PZT) and have 200 µm and 100 µm thickness respectively. For the construction of the A-CU, the transducer is bonded on the Si crystal using a thin film of vacuum epoxy, ensuring low acoustic losses at the transducer-crystal interface. After bonding to the Si crystal and manufacturing of the necessary electrical driving circuit, the system acoustic impedance resonant peaks are measured via a broadband network analyzer (Agilent E5061A, 300kHz - 1.5GHz). It was found that the peaks are shifted to 10.92 MHz and 21.4 MHz respectively. In the experiments the devices are driven in these measured frequencies to optimize acoustic pressure generation; however, for simplicity we will refer to them in the text using their nominal frequencies. For the electrical driving of the transducer, a ~200 mW, 10 Vpp signal generator delivering sinusoidal signals up to 60 MHz is used. For the 20 MHz prototype, an additional impedance matching circuit is developed to improve electrical coupling of the prototype to the output of the signal generator.

### C. Computational model

As mentioned earlier, the light intensity distribution at the CCD camera can be used to estimate the acoustic pressure distribution inside the crystals via a specially developed computational model. In the model, a crystal with height $L_y$ and distance from the imaging screen (camera sensor) $L_x$ is assumed. The angle of refraction $\theta_r$ of a single light ray after propagation inside the crystal is given by[42]:

$$\tan \theta_r = -\frac{C_{n0}}{n_0}\frac{dp}{dy}w \qquad (1)$$

where $C_{n0}$ is the piezo-optic coefficient of the crystal, $n_0$ is the refractive index in the absence of acoustic excitation, $dp/dy$ is the pressure gradient and $w$ is the optical path in the crystal.

Let $y$ be the vertical position on the crystal sample and $y_s$ the vertical position on the imaging screen. The divergence $\Delta y = y_s - y$ of a light ray propagating through an acoustically modulated crystal is given by:

$$\Delta y = \tan \theta_r \, L_x \tag{2}$$

so that:

$$y_s = y - \frac{C_{n0}}{n_0} \frac{dp}{dy} w L_x \tag{3}$$

A perfectly harmonic pressure $p(y)$ along the $y$ axis of the crystal can be expressed as:

$$p(y) = p_0 \sin(ky) \tag{4}$$

where $k$ is the spatial angular frequency of the acoustic wave, while the pressure gradient is given by:

$$\frac{dp}{dy} = p_0 k \cos(ky) \tag{5}$$

By plugging Eq. (3) in Eq. (5) we get:

$$y_s = y - k \frac{C_{n0} p_0}{n_0} w L_x \cos(ky) \tag{6}$$

The intensity $I_s(y_s)$ observed at a position $y_s$ on the screen is the result of deflection of light rays from different points of the crystal, due to the angular divergence caused by the pressure-induced modulation of the crystal's refractive index. In the real experiment, the screen is discretized by the camera pixels, hence it consists of elementary surfaces with the size $L_p$ of a pixel. In the 2D problem considered here, a range of values $[y_1, y_2]$ on the screen lie within the same pixel. So, a single pixel $y_p$ at the discretized screen domain corresponds to $y_s \in [y_1, y_2]$. The intensity on the pixel $y_p$ is given by:

$$I(y_p) = \int_{y_1}^{y_2} I_{y_s} dy_s \tag{7}$$

For the case of harmonic modulation, the equation $y = f(y_s)$ is transcendental and cannot be solved analytically. Hence, a numerical approach is adopted. Let us first assume homogeneous laser intensity on the crystal surface, so that:

$$I(y) = \text{const}, y \in L_y \tag{8}$$

The vertical dimension of the crystal sample at the laser beam entrance is discretized, e.g., with steps of 10 nm. The sufficiently fine discretization for each scenario can be easily identified by progressively refining the step size up to the point that the final prediction of the model does not change further. Each light ray entering the sample at a position $y$ is deflected onto a position $y_s$ on the imaging screen according to Eq. (6). The distribution of rays on the screen positions $y_s$ is grouped into pixels $y_p$ by taking the histogram:

$$I_p(y_p) = \text{hist}(y_s, N_p) \tag{9}$$

where $\text{hist}(a, b)$ denotes the histogram of a signal $a$ with a number of bars $b$ and $N_p = L_y/L_p$ is the number of pixels on the camera. In this way, the width of each bar in the histogram equals the pixel size.

In case of an inhomogeneous distribution of the initial radiation intensity $I(y)$, the intensity distribution $I_s(y_s)$ on the camera plane is found via the Jacobian transformation. Although beam expansion can lead to a nearly homogeneous intensity distribution, this scenario is very common when probing with a laser beam, which typically has a Gaussian-like radial intensity distribution. Now, assuming there is no light lost due to refraction, all the initial optical intensity reaches the camera sensor, so that:

$$\int_{L_y} I(y)dy = \int_{L_y} I_s(y_s) \, dy_s \tag{10}$$

from which we get:

$$I_s(y_s) = \frac{I(y)}{\frac{dy_s}{dy}} \tag{11}$$

The derivative $\frac{dy_s}{dy}$ can be calculated from Eq.6:

$$\frac{dy_s}{dy} = 1 + k^2 \frac{C_{n0}p_0}{n_0} wL_x \sin(ky) \tag{12}$$

Finally, by substituting Eq. (12) into Eq. (11), we derive the intensity $I_s$ at $y_s$.

$$I_s(y_s) = \frac{I(y)}{1 + k^2 \frac{C_{n0}p_0}{n_0} wL_x \sin(ky)} \tag{13}$$

Eq. (13) links the intensity $I_s(y_s)$ captured by the camera to the pressure amplitude $p_0$ of the acoustic wave.

It should be mentioned that, for the method to provide consistent results, the distance $L_x$ should be large enough so that the divergence $\Delta y$ causes a detectable intensity redistribution at the camera, while at the same time, $L_x$ should not be too large to avoid crossings and interference of the deflected rays, which could compromise the acoustic field estimation accuracy. The next subsection describes the signal processing procedure for extracting quantitative information from the measurements regarding the acoustic wavelength, acoustic pressure and lattice displacement induced in the crystal by the piezoelectric excitation.

### A. Signal analysis

To quantify the 2D spatial distribution of the acoustic pressure field from the images of the light intensity distribution the following processing steps are applied. To clearly reveal the intensity distribution of the laser beam due to the pressure gradients in the crystal sample, the modulated image (signal) is divided by the unmodulated image (reference), as dictated by Eq. (13) from which we get:

$$\frac{I_s(y_s)}{I(y)} = I_{\text{div}}(y_s) = \frac{1}{1 + k^2 \frac{C_{n0} p_0}{n_0} w L_x \sin(ky)} \tag{14}$$

The division produces a new image (divided), in which the impact of any irregularities of the laser beam profile on the observed intensity distribution is suppressed. Then, the 1D representation of the intensity modulation in the $y$ direction is extracted by taking a lineout on the divided image. An example of this procedure can be seen in FIG. 3, where the laser intensity distribution is shown for the test device (see next section) driven at 25 MHz with $10\,\text{V}_{\text{pp}}$. In FIG. 3(a), the red dashed line corresponds to the experimentally evaluated intensity along the y axis, as extracted by the divided image lineout illustrated in FIG. 3(b). The signals are normalized so that the unperturbed intensity level takes the value 1. Respectively, the solid blue line corresponds to the model estimation, where $p_0$ is calculated from Eq. 14. For the aforementioned case of the test device, the acoustic pressure is estimated to be $p_0 = 80 \pm 5\,\text{kPa}$, with a wavelength of 340±20 μm. Finally, the displacement $\Delta l_c$ of the lattice planes can also be estimated from the pressure $p_0$, using the known formula:

$$\Delta l_c = \frac{p_0}{\rho v \omega} \tag{15}$$

where $\rho, v$ are the mass density and the speed of sound in the crystal, respectively, while $\omega$ is the angular frequency of the acoustic wave.

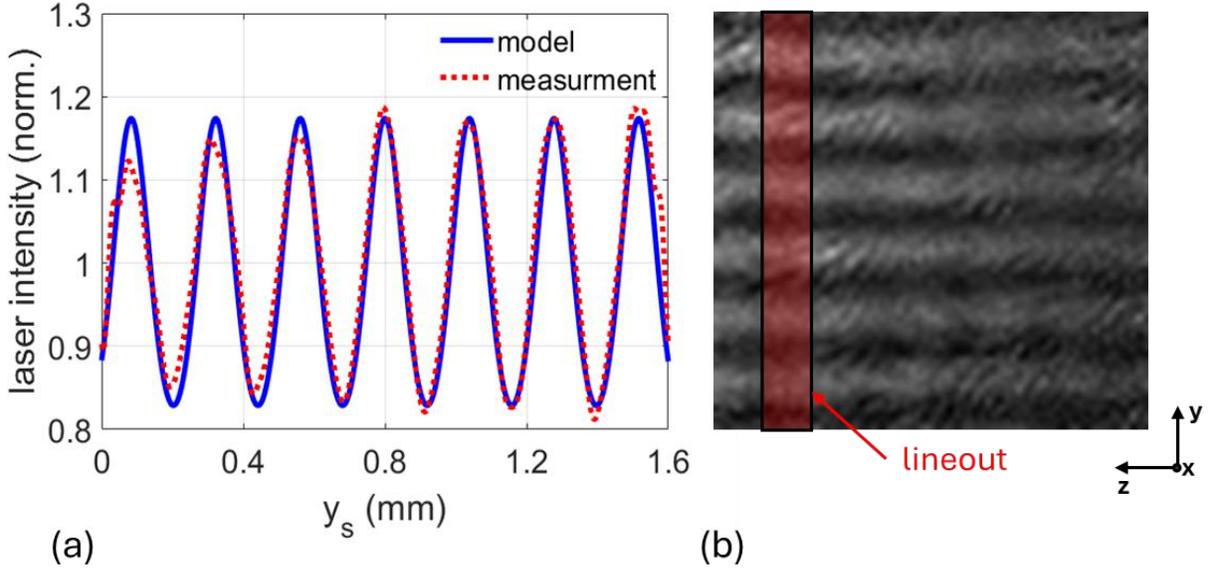

FIG. 3. (a) Modeled (solid blue line) and lineout of the measured laser intensity distribution (dashed red line) for the test device excited by 25 MHz and 10 Vpp. (b) Captured image of the redistributed laser intensity due to the travelling AW. The red marker illustrates the lineout region.

## III.   RESULTS

### A.   Model validation

To validate the computational model, a commercial acousto-optic modulator test device (Gooch & Housego, QS41-4C-B) based on a fused silica (FS) crystal was used. The device has a nominal operation frequency of 40.65 MHz but can be driven at a broad frequency range from 20 MHz to 60 MHz. Also, thanks to the long optical path of the FS crystal (5 cm), the measurement sensitivity is high so that acoustic waves can be detected with driving voltages <5 $V_{pp}$. These properties are exploited here to confirm the consistency of the model's predictions by calculating the acoustic pressure for varying:

a) driving voltage $V \in [5\ V_{pp}, 10\ V_{pp}]$.

b) driving frequency $f \in [20\ \text{MHz}, 60\ \text{MHz}]$.

c) distance of the sample from the camera sensor $L_x = 3$ cm and $6$ cm.

For a piezoelectric transducer driven in the linear regime at a specific frequency, the relationship between the driving voltage $V$ and the displacement $\Delta l_{\text{piezo}}$ of the transducer surface is approximately linear and is determined by the piezoelectric coefficient:

$$C_{\text{piezo}} = \frac{\Delta l_{\text{piezo}}}{V} \tag{16}$$

Since the pressure $p_0$ is linearly related to $\Delta l_{\text{piezo}}$, the pressure / voltage relationship is also expected to be linear. FIG. 4(a) presents the acoustic pressures and displacements estimated by the presented method for driving voltages from 5 to 10 $V_{\text{pp}}$. It can be seen that the expected linear relation is reproduced with high accuracy. The accuracy of the estimated values is supported by the theoretically expected displacements. For a typical LiNbO3 piezoelectric transducer with $C_{\text{LiNbO3}} = 6.2$ pm/V, a 10 Vpp driving voltage and total electrical and acoustic losses of ~30 % yield a ~40 pm displacement as estimated by our method (FIG. 4(a)), thus confirming the validity of the result.

Moreover, FIG. 4(b) shows the measured versus the calculated wavelengths of the acoustic waves inside the FS crystal for nine different driving frequencies in the range between 20 MHz and 60 MHz. The excellent matching of the experimental measurements and expected theoretical acoustic wavelengths proves that the detection method precisely captures the travelling multi-MHz acoustic waves.

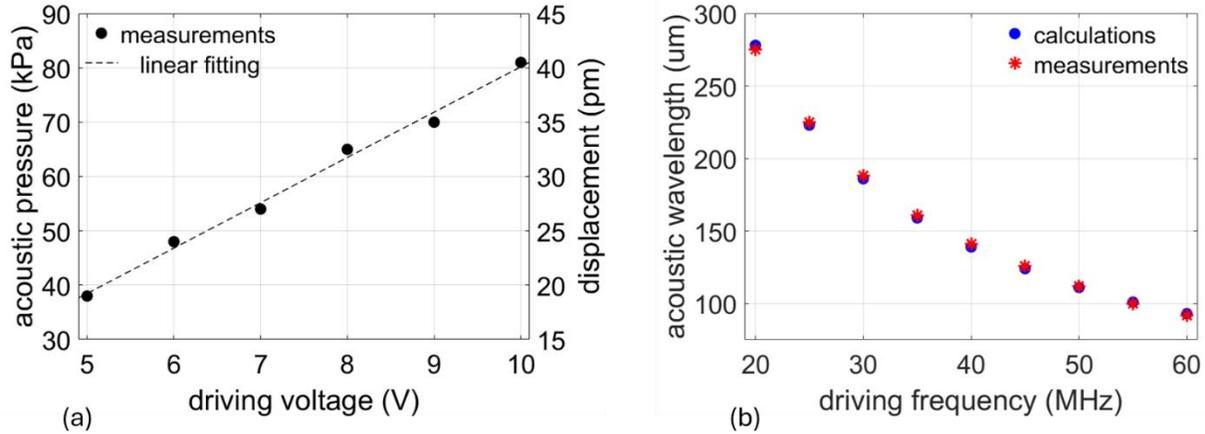

FIG. 4. (a) Experimentally derived acoustic pressure and lattice displacement (black dots) in the crystal of the test device driven at 25 MHz with different driving voltage amplitudes. The expected linearity is clearly demonstrated by the matching to the linear fitting curve. (b) Calculated (blue markers) and measured (red markers) wavelengths of the travelling AWs inside the test device for a range of driving frequencies, with a constant voltage amplitude of 10 Vpp. Again, the expected inverse proportionality is clearly depicted. In both graphs, the measurement error is of the order of the markers.

Finally, to test the method's robustness with respect to the distance between the crystal sample and the CDD camera, two different measurements, taken at $L_x = 3$ cm and $6$ cm, are analyzed. In this case, the dedicated driver of the test AOM device was used, with a fixed frequency of 40.65 MHz. For $L_x = 3$ cm, a pressure amplitude of $p_0 = 100 \ kPa$ was found, while for $L_x = 6$ cm, $p_0 = 95 \ kPa$. These results show that the distance $L_x$ does not significantly influence the acoustic pressure estimations ($\sim5\%$) proving the robustness of the proposed method. Note that this analysis is valid given that the distance $L_x$ is short enough so that no crossings of the deflected beams occur, as mentioned in section II.C.

**B. Characterization of the prototype devices**

FIG. 5 presents the results from the FLRI evaluation of the two prototype devices based on 1 mm-thick Si <100> plates driven at their measured resonant frequencies 10.92 MHz and 21.4 MHz, with a 10 Vpp driving voltage. Again, the measured and the numerically estimated laser intensity distribution in the $y_s$ axis are presented in FIG. 5(a) and (b), along with the original captured images (as insets). For the 10 MHz prototype, it can be seen that in the selected 2.1 mm observation window, the acoustic wave spans just over 3 periods. The obtained wavelength of the acoustic wave is 774±11 µm which is agreement with the theoretically expected (772 µm), considering the longitudinal speed of sound in silicon 8433 m/s. For the 20 MHz prototype, the number of periods increases to approximately 6 and the wavelength is 390±9 µm, which is also in agreement with the theoretically expected (394 µm).

The amplitude of the acoustic pressure for the 10 MHz and 20 MHz devices is measured to be 1530±17 kPa and 295±4 kPa, respectively, corresponding to displacement of the lattice planes of approximately 1.10 nm and 0.12 nm, respectively. The 1.10 nm displacement corresponds to twice the lattice constant of Si (0.543 nm), which is well within the range of interest for dynamic acoustic bending CUs[24]. Specifically, for the 10 MHz device, the combined parameters of 1.10 nm lattice displacement and 774 µm wavelength lie in the regime of intermediate amplitude and long period, which is suitable for generation of γ-rays by ultra-relativistic positron beams of several tens of GeV. For the 20 MHz device, the displacement of 0.12 nm, which is well below the Si lattice constant, combined with the 390 µm wavelength, classifies the device in the short amplitude long period (SALP) regime, which is also of interest for γ-ray generation from several tens of GeV positron beams. Such high quality positron beams, and particularly 20 GeV and 35 GeV beams, are currently available at CERN.

A critical aspect for the successful development of A-CUs is the spatial harmonicity of the acoustic pressure distribution inside the crystal. Non-harmonic components can have detrimental effects on the emitted γ radiation, as they can reduce photon yield in the energy of interest due to the generation of γ photons at undesirable energies. Moreover, anharmonicity can lead to a broadening of the generated γ-ray spectrum, hence depreciating the undulator's performance regarding narrowband γ-radiation emission. In this respect, the proposed characterization method is highly suitable, as it can image and quantify the acoustic field over a wide region of the crystal, hence allowing the evaluation of the harmonicity of the acoustic field across the region of the positron beam propagation. FIGS. 5(c) and 5(d) show the experimentally evaluated lattice modulations for the 10 MHz and 20 MHz devices, respectively, in the spatial frequency domain, as obtained by Fast Fourier Transform. It can be seen that the harmonic components at the targeted spatial frequencies 1295 m$^{-1}$ and 2538 m$^{-1}$, respectively, dominate the spatial frequency spectrum. Minor anharmonic spectral components appear, especially in the case of the 20 MHz device; however, their spectral energy is of the order of a few percent compared to that of the harmonic component. Considering also the results presented in Ref 17, where a relatively strong enhancement of undulation radiation was observed by simulation of an A-CU device with comparable anharmonicity, it becomes apparent that the devices presented here are very promising for the generation of narrowband γ-rays.

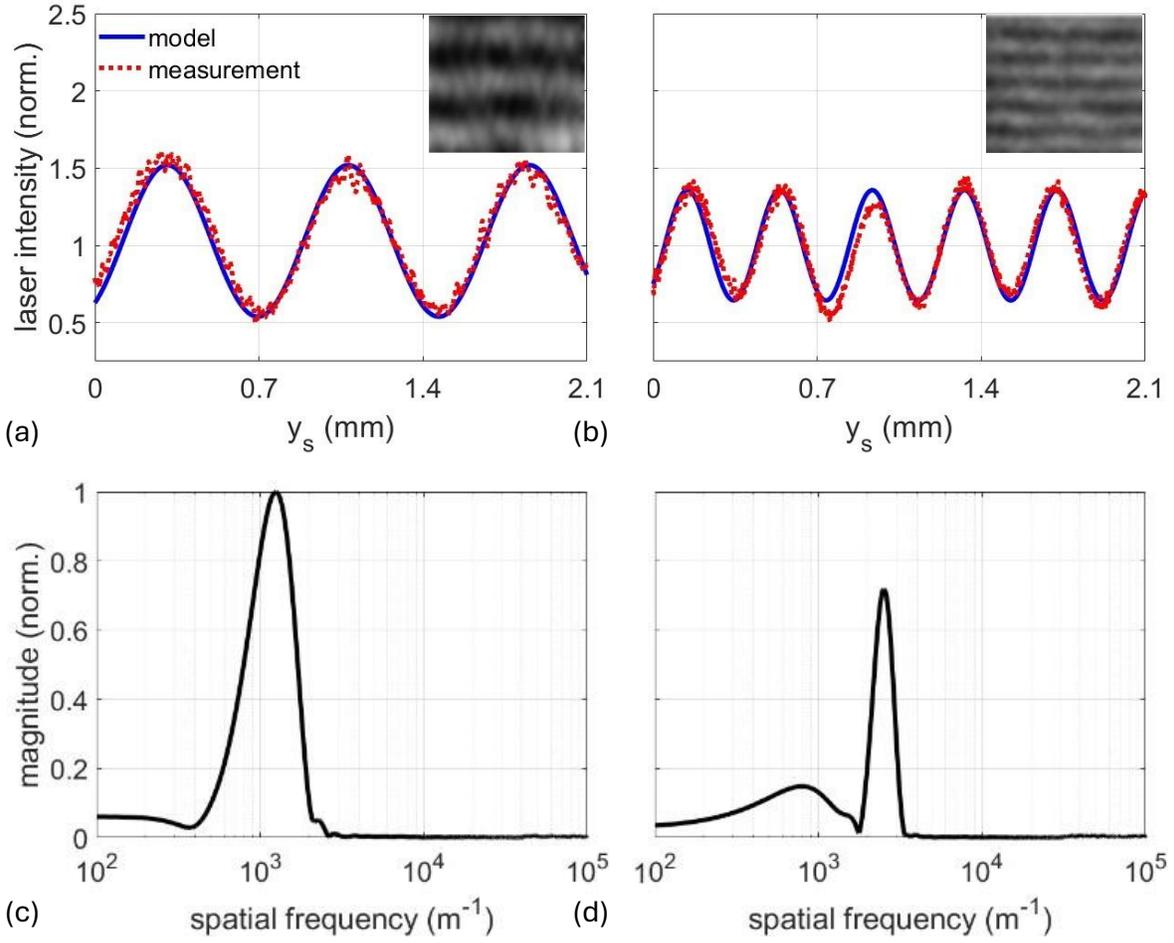

FIG. 5. (a-b) Modeled (solid blue line) and measured (dashed red line) spatial distribution (lineout) of the laser intensity after deflection inside acoustically modulated 1 mm thick Si crystals, driven at (a) 10.92 and (b) 21.4 MHz. The insets show the images captured by FLRI. (c-d) Spatial frequency spectra of the signals shown in (a-b), respectively, as obtained via FFT, demonstrating the harmonicity of the acoustic fields inside the crystals.

## IV. CONCLUSION

In this work, we presented a complete method for the precise characterization of traveling acoustic waves in the tens of MHz range, within crystalline materials. The method combines fast laser refraction imaging (FLRI) with a specially developed computational model that enables quantitative reconstruction of the acoustic field. Nanosecond laser pulses were used to image the traveling acoustic waves by capturing the deflection of light rays induced by pressure-dependent refractive index gradients transverse to the beam path. The resulting optical intensity patterns on the imaging screen were quantitatively mapped to acoustic pressure using the computational model. Validation of the method's estimations was achieved by use of a commercial acousto-optic modulator, driven in a range of different voltages and frequencies.

Here, experiments were carried out in 1 mm-thick Si<100> monocrystals, while acoustic excitation was applied by piezoelectric transducers operating at 10 MHz and 20 MHz, which were bonded to the silicon crystals. The very good agreement of the experimental estimations to the theoretically expected values clearly demonstrated the high sensitivity of the proposed method, proving that it can be used for imaging and quantification of acoustic fields even inside very thin crystal samples. Moreover, the method allows for imaging of a broad region of the sample with a single measurement, hence rendering it suitable for evaluation of the spatial distribution of the acoustic pressure in the crystal.

This work constitutes a first technological step toward the development of acoustically driven, crystal-based devices for generating tunable, directional, and narrowband γ-radiation via the undulation of ultra-relativistic charged particles. The induced harmonic bending of the crystal lattice can force channeled particles to follow near-sinusoidal trajectories, leading to emission of narrowband radiation in the range from hundreds of keV up to many MeV. The presented method can also be used for the development of other optoacoustic devices, for non-destructive testing of transparent media and characterization of ultrasonic transducers. Future work will focus on optimizing these devices for operation at higher frequencies and larger lattice deformations, as well as on conducting real γ-ray generation experiments at high-energy facilities.

## ACKNOWLEDGMENTS

The authors acknowledge financial support from the European Commission's Horizon Europe-EIC-Pathfinder-Open TECHNO-CLS (G.A. 101046458) project.

## AUTHOR DECLARATIONS

### Conflict of Interest

All authors declare no conflict of interest.

### DATA AVAILABILITY

The data related to this study are available from the corresponding authors upon reasonable request.